\documentclass[preprint, superscriptaddress, draft, amssymb, floats, byrevetex]{revtex4}
\usepackage{latexsym}
\usepackage{graphics}
\newcommand{\bmat}{\left(\begin{array}}
\newcommand{\emat}{\end{array}\right)}
\newcommand{\beq}{\begin{equation}}
\newcommand{\eeq}{\end{equation}}
\newcommand{\drawsquare}[2]{\hbox{%
\rule{#2pt}{#1pt}\hskip-#2pt%  left vertical
\rule{#1pt}{#2pt}\hskip-#1pt%  lower horizontal
\rule[#1pt]{#1pt}{#2pt}}\rule[#1pt]{#2pt}{#2pt}\hskip-#2pt%  upper horizontal
\rule{#2pt}{#1pt}}% right vertical

\newcommand{\fund}{\raisebox{-.5pt}{\drawsquare{6.5}{0.4}}}%  fund
\newcommand{\Ysymm}{\raisebox{-.5pt}{\drawsquare{6.5}{0.4}}\hskip-0.4pt%
        \raisebox{-.5pt}{\drawsquare{6.5}{0.4}}}%  symmetric second rank
\newcommand{\Yasymm}{\raisebox{-3.5pt}{\drawsquare{6.5}{0.4}}\hskip-6.9pt%
        \raisebox{3pt}{\drawsquare{6.5}{0.4}}}%  antisymmetric second rank
\newcommand{\antifund}{\overline{\fund}}

\def\yzero{\smash{\hbox{$y\kern-4pt\raise1pt\hbox{${}^\circ$}$}}}

\def\-{\hphantom{-}}
\def\ov{\overline}
\def\s2{\frac{1}{\sqrt2}}

\def\beq{\begin{equation}}
\def\eeq{\end{equation}}
\def\beqa{\begin{eqnarray}}
\def\eeqa{\end{eqnarray}}

\def\IF{\relax{\rm I\kern-.18em F}}
\def\II{\relax{\rm I\kern-.18em I}}
\def\IP{\relax{\rm I\kern-.18em P}}

\def\Dsl{\,\raise.15ex\hbox{/}\mkern-13.5mu D} %can be subscripted

\def\IC{\bf C}
\def\IZ{\bf Z}
\def\IT{\bf T}
\def\z2z2{$\IC^3/(\IZ_2\times\IZ_2)$}

\def\s{\sigma}

\def\z{\zeta}

\def\bo{{\raise-.3ex\hbox{\large$\Box$}}}               % D'Alembertian
\def\face{{\raise.2ex\hbox{$\displaystyle \bigodot$}\mskip-2.2mu \llap {$\ddot
        \smile$}}}                                      % happy face
\def\leftrightarrowfill{$\mathsurround=0pt \mathord\leftarrow \mkern-6mu
        \cleaders\hbox{$\mkern-2mu \mathord- \mkern-2mu$}\hfill
        \mkern-6mu \mathord\rightarrow$}       % <--> double differential
\def\dvec#1{\vbox{\ialign{##\crcr
        \leftrightarrowfill\crcr\noalign{\kern-1pt\nointerlineskip}
        $\hfil\displaystyle{#1}\hfil$\crcr}}}           % <--> accent
\def\beq{\begin{equation}}
\def\eeq{\end{equation}}

\def\beqx{\begin{displaymath}}
\def\eeqx{\end{displaymath}}

\def\beqa{\begin{eqnarray}}
\def\eeqa{\end{eqnarray}}
\def\NO{\nonumber}

\begin{document}
\title{
\normalsize \mbox{ }\hspace{\fill}
\begin{minipage}{12 cm}
{\tt UPR-1029-T, PUPT-2078, hep-th/0303197}{\hfill}
\end{minipage}\\[5ex]
{\large\bf More Supersymmetric Standard-like  Models from
Intersecting D6-branes on Type IIA Orientifolds
\\[1ex]}}
\date{\today}
\author{Mirjam Cveti\v c}
\affiliation{School of Natural Sciences, Institute for Advanced
Studies, Princeton NJ 08540 USA
and\\
 Department of Physics and Astronomy, Rutgers University,
Piscataway, NJ 08855-0849 USA} \altaffiliation{On Sabbatic Leave
from the University of Pennsylvania}
\author{Ioannis Papadimitriou}
\affiliation{Princeton University, Princeton NJ 08540 USA}
\altaffiliation{Exchange Scholar from the University of
Pennsylvania}

\begin{abstract}We present  new classes of  supersymmetric
Standard-like models from  type IIA $\IT^6/(\IZ_2\times \IZ_2)$
orientifold with intersecting D6-branes. D6-branes can wrap
general  supersymmetric three-cycles of $\IT^6=\IT^2\times
\IT^2\times \IT^2$, and any  $\IT^2$ is allowed to be tilted. The
models still suffer from additional exotics, however  we obtained
solutions with fewer Higgs doublets,  as well as models with all
three families of  left-handed quarks and leptons arising from the
same intersecting  sector, and  examples of a genuine left-right
symmetric model  with three copies of  left-handed and
right-handed families of quarks  and leptons.
\end{abstract} \maketitle

\section{Introduction}

The intersecting  D-brane configurations of  Type II string
theory, compactified on orientifolds  play an important role in
the construction of four-dimensional solutions
\cite{bgkl,afiru,bkl,imr,magnetised}. In particular the appearance
of the chiral matter \cite{bdl,bachas}  at the brane intersection
provides a promising starting point to construct the models  with
potentially quasi-realistic particle physics features.

Techniques to  construction of intersecting D-brane models
 were  developed,  primarily for
non-supersymmetric constructions, in
\cite{bgkl,afiru,bkl,imr,magnetised} (and subsequently explored in
\cite{bonn,bklo,bailin,kokorelis}). Many non-supersymmetric models
with quasi-realistic features of the Standard-like and
grand-unified models can be obtained, satisfying the Ramond-Ramond
tadpole cancellation conditions. However, since the models are
non-supersymmetric, there are uncancelled
Neveu-Schwarz-Neveu-Schwarz tadpoles as well as the radiative
corrections at the effective theory level, which are of the string
scale.  Since the intersecting D6-branes typically have no common
transverse in the internal space, the string scale is  of the
Planck scale. Therefore the tree-level constructions typically
suffer from large Planck scale corrections at the loop level.
 %(However  examples \cite{CIMD5,Kok2} with
%intersecting D5-branes have been given, where the string scale can
%be as low as the TeV scale.)

On the other hand,  constructions of supersymmetric models
 turn out to
be extremely constraining.  The first supersymmetric model with
intersecting D6-branes  and the features of the supersymmetric
Standard-like models has been constructed \cite{CSU1,CSU2,CSU3}.
They are based on the $\IZ_2\times \IZ_2$ orientifold   with
D6-branes wrapping  specific supersymmetric three-cycles of the
six-torus.
 Interestingly, the embedding
of supersymmetric four-dimensional models with intersecting
D6-branes has a lift \cite{CSU2,CSU3} into M-theory that
corresponds to the compactification of M-theory on a singular
$G_2$  holonomy manifold \cite{CSU1,CSU2,AW,Witten,aW}.

Within the above class of models   examples of supersymmetric
four-family \cite{CSU2} and three-family \cite{CPS} SU(5) models
were also constructed. In particular,  in \cite{CPS} a systematic
analysis for three-family models with D6-branes wrapping general
three-cycles of internal six-torus and for general tilted two-tori
was  given. There are of the order of 80 models in this class of
constructions and they all  necessarily contain three-copies of
$\bf 15$ of SU(5).

Recently three-family supersymmetric left-right symmetric models
based on $\IZ_4$  \cite{blumrecent} and  $\IZ_4\times \IZ_2$
\cite{Honecker} orientifolds with  intersecting D6-branes were
obtained. In these models the left-right symmetric gauge structure
was obtained via  brane recombinations and thus the final model
does not have an explicit toroidal orientifold construction, where
conformal  field theory techniques could be applied for the
calculation of the full spectrum and couplings.

 While providing an explicit example of
 Standard-like model gauge group and three
copies of quarks and leptons, the original  construction
\cite{CSU1,CSU2}
 suffers from a number of
phenomenological difficulties. One is   the existence of the
adjoint matter on each brane, which is a  general property
supersymmetric as well as  non-supersymmetric toroidal orientifold
constructions. It is due to the fact that the typical cycles of
the tori, wrapped by D6-branes,  are not rigid. Addressing
Calabi-Yau compactifications with rigid supersymmetric cycles
would therefore be one avenue to pursue, however the calculational
techniques of conformal field theory may not be applicable there.
The original  construction  also contains charged exotics, some of
them fractionally charged,  and it has many Standard-Model Higgs
doublets (12 $H_U$ and 12 $H_D$). The model also has two
additional non-anomalous  U(1) factors, and for one of them there
are no light elementary Higgs field candidates to spontaneously
break it. (See \cite{CLS1} for details.)

The model, however, has an additional quasi-hidden gauge sector
that is confining. This sector may allow for  binding of exotics
into composite states that have charge assignments of the (fourth)
left-handed family, as discussed in \cite{CLS1}. Another
interesting implication of this confining sector is a possibility
of  gaugino condensation which may trigger supersymmetry breaking
and stabilization  of moduli  \cite{CLW}.

Phenomenology of both non-supersymmetric \cite{imr,kokorelis,CIM1}
and supersymmetric \cite{CLS1,CLS2,CLW} models was addressed,
however the calculations of couplings has been only recently
studied. The tree-level gauge couplings and their low energy
implications have been studied (See e.g., \cite{CLS1} and
references therein). A calculation of gauge coupling threshold
corrections \cite{Lust} is also of  interest, since it could be
compared to the strong coupling limit of M-theory compactified on
the corresponding $G_2$ holonomy space \cite{FW}.

Most recently, progress has been made for  the calculations of
couplings for states at the intersections. In particular the
Yukawa couplings of quarks and leptons to Higgs fields is of most
interest. These couplings are expected to obtain contributions
from the world-sheet (disk) instantons associated with the open
strings stretching between the three intersections \cite{afiru}.
Recently, the complete instanton sum for toroidal orientifolds has
been addressed in  \cite{CIM}.  The full coupling, that includes
the classical and the quantum part contribution has been
calculated in \cite{CP}, employing the conformal field theory
technique. (See also \cite{AO} for a conformal field theory
calculation of the classical part of the Yukawa couplings.)

The purpose of this paper is to present new classes of
supersymmetric Standard-like  models within $ {\IT^6/( \IZ_2\times
\IZ_2)}$ orientifolds with intersecting D6-branes. We generalize
the original constructions \cite{CSU1,CSU2} by allowing for
D6-branes wrapping most general three-cycles, constructed as a
product of one-cycles on each two-torus, and for any of the
two-tori to be tilted. The aim is to demonstrate existence of
Standard-like models with fewer Higgs doublets and fewer exotics.
As a by-product we also obtain the models that have three-families
of left-handed quarks and leptons arising for the same
intersecting sector, as well as the models where the number of the
right-handed families is also three. [In the original construction
one  of the families arises from a different intersection sector.
There the number of right-handed families was even, which in turn
required additional exotics to cancel the Standard-Model gauge
group anomalies.]

The models that we constructed are basically  descendants of the
left-right symmetric models, i.e. the  non-anomalous hypercharge
is obtained as a linear combination of  non-anomalous U(1) charges
that lie within the Pati-Salam left-right gauge group structure
\cite{PatiSalam}. As a consequence we typically obtain  two
additional non-anomalous U(1)  gauge group factors in the
observable sector, just as in the original construction.

% (It would be interesting to generalize these constructions
%to cases where the hypercharge would not be a descendant of the
%left-right symmetric model.)

The paper is organized as follows. In Section II we summarize the features
of the construction. In Section III we  give details about
specific classes of new models. Conclusions and open avenues are given in
Section IV.

\section{Construction of
Supersymmetric Models
%Left Right Models
from a $\IT^6 /\IZ_2 \times \IZ_2$ Orientifold with Intersecting
D6-branes
}

Let us, for completeness, present the orientifold construction in
\cite{CSU2} as well as some minor changes in notation which were
introduced in \cite{CPS}. The starting point is type IIA string
theory compactified on a $\IT^6/(\IZ_2\times \IZ_2)$ orientifold.
We take $\IT^6=\IT^2\times\IT^2\times\IT^2$ to be the product of
three 2-tori and introduce complex coordinates $z_i$, $i=1,\; 2,\;
3$ parameterizing each of the 2-tori. The orbifold group
generators $\theta$, $\omega$ then act on  $\IT^6$ as \beqa
& \theta: & (z_1,z_2,z_3) \to (-z_1,-z_2,z_3) \nonumber \\
& \omega: & (z_1,z_2,z_3) \to (z_1,-z_2,-z_3).
\label{orbifold}\eeqa The orientifold projection is implemented by
gauging the symmetry $\Omega R$, where $\Omega$ is world-sheet
parity, and $R$ acts as \beqa R: (z_1,z_2,z_3) \to ({\ov z}_1,{\ov
z}_2,{\ov z}_3). \label{orientifold}\eeqa There are then  four
kinds of orientifold 6-planes (O6-planes), associated with the
actions of $\Omega R$, $\Omega R\theta$, $\Omega R \omega$, and
$\Omega R\theta\omega$ respectively. To cancel the RR charge of
the O6-planes, D6-branes wrapped on factorized three-cycles are
introduced. We consider $K$ stacks of $N_a$ D6-branes,
$a=1,\ldots, K$, wrapped on the $n_a^i[a_i]+m_a^i[b_i]$ cycle in
the $i^{th}$ two-torus. There are only two choices of the complex
structure consistent with the orientifold projection as explained
in \cite{CPS}. In either case a generic two cycle is labelled by
$(n_a^i,l_a^i)$, where in terms of the wrapping numbers
$l_{a}^{i}\equiv m_{a}^{i}$ for a rectangular torus and
$l_{a}^{i}\equiv 2\tilde{m}_{a}^{i}=2m_{a}^{i}+n_{a}^{i}$ for a
tilted torus. For a stack of $N_a$ D6-branes along cycle
$(n_a^i,l_a^i)$ we also need to include their $\Omega R$ images
$N_{a'}$ with wrapping numbers $(n_a^i,-l_a^i)$. For branes on top
of the O6-planes we also count branes and their images
independently. So the homology three-cycles for stack $a$ of $N_a$
D6-branes and its orientifold image $a'$ take the form \beq
[\Pi_a]=\prod_{i=1}^{3}\left(n_{a}^{i}[a_i]+2^{-\beta_i}l_{a}^{i}[b_i]\right),\;\;\;
\left[\Pi_{a'}\right]=\prod_{i=1}^{3}\left(n_{a}^{i}[a_i]-2^{-\beta_i}l_{a}^{i}[b_i]\right)
\eeq where $\beta_i=0$ if the $i$th torus is not tilted and
$\beta_i=1$ if it is tilted.  In particular, the homology
three-cycles wrapped by the four orientifold planes are (our
normalization of the orientifold cycles, explained in more detail
in \cite{CPS}, is slightly different from the one in \cite{CSU2}.
Here we include the multiplicity of the orientifold planes which
allows for a uniform treatment of rectangular and tilted tori.)
\beq
\begin{array}{clcl}
\Omega R: & [\Pi_1]=8[a_1]\times[a_2]\times[a_3],& \Omega R\omega:
&
[\Pi_2]=-2^{3-\beta_2-\beta_3}[a_1]\times[b_2]\times[b_3]\\
\\ \Omega R\theta\omega: &
[\Pi_3]=-2^{3-\beta_1-\beta_3}[b_1]\times[a_2]\times[b_3], &
\Omega R\theta: &
[\Pi_4]=-2^{3-\beta_1-\beta_2}[b_1]\times[b_2]\times[a_3]
\end{array}\label{orienticycles}\eeq
If we define $[\Pi_{O6}]=[\Pi_1]+[\Pi_2]+[\Pi_3]+[\Pi_4]$, the
chiral spectrum in the open string sector can be described in
terms of the intersection numbers
\beq
\begin{array}{ll}
I_{ab}=[\Pi_a][\Pi_b]=2^{-k}\prod_{i=1}^3(n_a^il_b^i-n_b^il_a^i),&
I_{ab'}=[\Pi_a]\left[\Pi_{b'}\right]=-2^{-k}\prod_{i=1}^3(n_{a}^il_b^i+n_b^il_a^i)
\\\\
I_{aa'}=[\Pi_a]\left[\Pi_{a'}\right]=-2^{3-k}\prod_{i=1}^3(n_a^il_a^i),&
\\\\
\multicolumn{2}{l}{I_{aO6}=[\Pi_a][\Pi_{O6}]=2^{3-k}(-l_a^1l_a^2l_a^3+l_a^1n_a^2n_a^3+n_a^1l_a^2n_a^3+n_a^1n_a^2l_a^3)}
\end{array}
\label{intersections}\eeq where $k=\beta_1+\beta_2+\beta_3$ is the
total number of tilted tori.

The model is constrained by the tadpole cancellation conditions
\beq \sum_a N_a [\Pi_a]+\sum_a N_a
\left[\Pi_{a'}\right]-4[\Pi_{O6}]=0 \eeq as well as the conditions
to preserve ${\cal N}=1$ supersymmetry in $D=4$ \cite{bdl}, namely
each stack of $D6$-branes should be related to the $\Omega
R$-plane by an $SU(3)$ rotation. If $\theta_i$ is the angle the
$D6$-brane makes with the $\Omega R$-plane in the $i$th torus,
then supersymmetry requires $\theta_1+\theta_2+\theta_3=0\;\; {\rm
mod}\;\; 2\pi$. In \cite{CPS} it was shown that this condition can
be rewritten as \begin{eqnarray} -x_A l_a^1l_a^2l_a^3+x_B
l_a^1n_a^2n_a^3+x_C n_a^1l_a^2n_a^3+x_D n_a^1n_a^2l_a^3=0 \nonumber\\\nonumber \\
-n_a^1n_a^2n_a^3/x_A+n_a^1l_a^2l_a^3/x_B+l_a^1n_a^2l_a^3/x_C+l_a^1l_a^2n_a^3/x_D<0
\label{susyconditions}
\end{eqnarray} where $x_A=\lambda,\;
x_B=\lambda 2^{\beta_2+\beta3}/\chi_2\chi_3,\; x_C=\lambda
2^{\beta_1+\beta3}/\chi_1\chi_3,\; x_D=\lambda
2^{\beta_1+\beta2}/\chi_1\chi_2$ and $\chi_i=(R_2/R_1)_i$ are the
complex structure moduli. $\lambda$ is a positive parameter
without physical significance.

The open string spectrum of these constructions for branes at
generic angles was discussed in detail in \cite{CSU2}. We
summarize the results in table~\ref{matter}. In particular, the
$aa$ sector arising from open strings stretching within a single
stack of D6$_a$-branes contains $U(N_a/2)$ gauge fields as well as
three adjoint $N=1$ chiral multiplets which are the moduli
associated with the non-rigidness of the three-cycles the
D6-branes wrap. The $ab+ba$ sector contains $I_{ab}$ chiral
multiplets  in the $(\fund_a,\antifund_b)$ representation of
$U(N_a/2)\times U(N_b/2)$ while $ab'+b'a$ contains $I_{ab'}$
chiral multiplets in the bifundamental $(\fund_a,\fund_b)$.
Finally, the $aa'+a'a$ sector contains symmetric and antisymmetric
representations of the $U(N_a/2)$ gauge group with multiplicities
given in table~\ref{matter}. Notice that because of the change of
notation from \cite{CSU1,CSU2}, the multiplicities of $\Ysymm$ and
$\Yasymm$ have slightly different expressions here.
\begin{table}
[htb] \footnotesize
\renewcommand{\arraystretch}{1.25}
\begin{center}
\begin{tabular}{|c|c|}
\hline {\bf Sector} & \phantom{more space inside this box}{\bf
Representation}
\phantom{more space inside this box} \\
\hline\hline
$aa$   & $U(N_a/2)$ vector multiplet  \\
       & 3 Adj. chiral multiplets  \\
\hline
$ab+ba$   & $I_{ab}$ $(\fund_a,\antifund_b)$ fermions   \\
\hline
$ab'+b'a$ & $I_{ab'}$ $(\fund_a,\fund_b)$ fermions \\
\hline $aa'+a'a$ &$-\frac 12 (I_{aa'} - \frac 12 I_{a,O6})\;\;
\Ysymm\;\;$ fermions \\
          & $-\frac 12 (I_{aa'} + \frac 12 I_{a,O6}) \;\;
\Yasymm\;\;$ fermions \\
\hline
\end{tabular}
\end{center}
\caption{\small General spectrum on D6-branes at generic angles,
i.e., angles that are not parallel to any O6-plane in all three
tori. The spectrum is valid for both tilted and untilted tori. The
models may contain additional non-chiral pieces in the $aa'$
sector and in $ab$, $ab'$ sectors with zero intersection, if the
relevant branes overlap. In supersymmetric situations, scalars
combine with the fermions given above to form chiral
supermultiplets. \label{matter} }
\end{table}

\section{Features of the  Supersymmetric Standard-like models}

In trying to solve the tadpole and supersymmetry conditions, along
with other phenomenological constraints such as three families of
quarks and leptons and semi realistic gauge groups, it is very
useful to classify all possible supersymmetric brane
configurations that can be used in any model based on this
construction. This is indeed possible because supersymmetry, in
contrast to the tadpole cancellation conditions, constrains each
stack of branes individually. In \cite{CPS} we have classified all
supersymmetric brane configurations in the context of the
$\IT^6/(\IZ_2\times\IZ_2)$ orientifold described above. We found
that a brane wrapping a 3-cycle which has three wrapping numbers
$n_i,\; l_i$ equal to zero is necessarily parallel to one of the
four orientifold planes. Such brane configurations (filler branes)
trivially satisfy the supersymmetry conditions and so can help
solve the tadpole conditions. There can be no supersymmetric brane
configuration along a cycle with two vanishing wrapping numbers
and hence the only other possibilities are one or no zero wrapping
numbers. Each of these gives a supersymmetric configuration
together with a constraint for the complex structure moduli
\cite{CPS}. Since there are three independent moduli parameters we
can generically include only up to three such brane
configurations. Otherwise the system is overconstrained. Having
these building blocks at our disposal we can start looking for
models with the desired gauge group and number of families.

\begin{table}
[htb] \footnotesize
\renewcommand{\arraystretch}{1.0}
\begin{center}
\begin{tabular}{|c||c|c||c|c|c|c|c|c|c|}
\hline
    \rm{model} I.1 & \multicolumn{9}{c|}{\phantom{more space}$U(3)\times U(1)\times U(2)\times USp(8)\times USp(2)\times USp(4)$\phantom{more space}}\\
\hline\hline \rm{stack} & $N$ & $(n^1,l^1)\times (n^2,l^2)\times
(n^3,l^3)$ & \phantom{2 }$n_{\Ysymm}$\phantom{2 } & \phantom{2 }$n_{\Yasymm}$\phantom{2 }  & \phantom{2 }$b$\phantom{2 } & \phantom{2 }$b'$\phantom{2 } & \phantom{2 }1 \phantom{2 }& \phantom{2 }2 \phantom{2 }& \phantom{2 }3\phantom{2 }\\
\hline
    $a$&  6+2& $(0,1)\times (1,-1)\times (1,-1)$ & 0 & 0  & 0  & 3 & -1 & 1 & 0 \\
    $b$&   4& $(3,-2)\times (0,1)\times (1,-1)$ & 1 & -1  &-  &- & -2& 0 & 3\\
\hline
    1&   8& $(1,0)\times (1,0)\times (2,0)$ & \multicolumn{7}{c|}{$2x_A=2x_B=3x_C$}\\
    2&   2& $(1,0)\times (0,1)\times (0,-2)$& \multicolumn{7}{c|}{}\\
    3&   4& $(0,1)\times (1,0)\times (0,-2)$ & \multicolumn{7}{c|}{}\\
\hline
\end{tabular}
%\end{center}
%\caption{Model I.1} \label{modelI.1}
%\end{table}

%\begin{table}
%[htb] \footnotesize
%\renewcommand{\arraystretch}{1.0}
%\begin{center}
\begin{tabular}{|c||c|c||c|c|c|c|c|c|c|}
\hline
    \rm{model} I.2 & \multicolumn{9}{c|}{\phantom{more space} $U(3)\times U(1)\times U(2)\times USp(8)\times USp(2)\times USp(4)$\phantom{more space}}\\
\hline\hline \rm{stack} & $N$ & $(n^1,l^1)\times (n^2,l^2)\times
(n^3,l^3)$ & \phantom{2 }$n_{\Ysymm}$\phantom{2 } &\phantom{2 } $n_{\Yasymm}$\phantom{2 }  & \phantom{2 }$b$ \phantom{2 }& \phantom{2 }$b'$\phantom{2 }  & \phantom{2 }1\phantom{2 } & \phantom{2 }2 \phantom{2 }& \phantom{2}3 \phantom{2}\\
\hline
    $a$&  6+2& $(0,1)\times (1,-1)\times (1,-1)$ & 0 & 0  & 1  & 2 & -1 & 1 & 0 \\
    $b$&  4& $(-1,2)\times (0,1)\times (-1,3)$ & -5 & 5  &-  &- & -6 & 0 & 1  \\
\hline
    1&   8& $(1,0)\times (1,0)\times (2,0)$ & \multicolumn{7}{c|}{$6x_A=6x_B=x_C$}\\
    2&   2& $(1,0)\times (0,1)\times (0,-2)$& \multicolumn{7}{c|}{}\\
    3&   4& $(0,1)\times (1,0)\times (0,-2)$ & \multicolumn{7}{c|}{}\\
\hline
\end{tabular}
\end{center}
\caption{D6-brane configurations and intersection numbers for the
three-family Standard-like models (continued in table
\ref{models2}). Here $a, b$ denote the stacks of D-branes not
parallel to the orientifold planes, giving $U(N_a/2)$ gauge group,
while $a', b'$ denote their $\Omega R$ image. 1,2,3,4 denote
filler branes respectively along the $\Omega R$, $\Omega R\omega$,
$\Omega R\theta\omega$ and $\Omega R\theta$ orientifold plane,
resulting in a $USp(N^{(i)})$ gauge group. $N$ is the number of
branes in each stack. The third column shows the wrapping number
of the various branes. In all models only the third torus has a
non-trivial complex structure. The intersection numbers between
the various stacks are given in the remaining columns to the
right. For example, the intersection number $I_{ac}$ between
stacks $a$ and $b$ is found in row $a$ column $b$. For convenience
we also list the relation among the moduli imposed by the
supersymmetry conditions, as well as the gauge group for each
model.} \label{models}
\end{table}

\begin{table}
[htb] \footnotesize
\renewcommand{\arraystretch}{1.0}
\begin{center}
\begin{tabular}{|c||c|c||c|c|c|c|c|c|c|c|}
\hline
    \rm{model} I.3 & \multicolumn{10}{c|}{\phantom{morespace}$U(3)\times U(1)\times U(2)\times USp(8)\times USp(2)\times USp(4)
    \times USp(2)$\phantom{morespac}}\\
\hline\hline \rm{stack} & $N$ & $(n^1,l^1)\times (n^2,l^2)\times
(n^3,l^3)$ & \phantom{2}$n_{\Ysymm}$ \phantom{2}&\phantom{2} $n_{\Yasymm}$ \phantom{2} & \phantom{2}$b$\phantom{2} & \phantom{2}$b'$ \phantom{2 } &\phantom{2}1\phantom{2} & \phantom{2}2\phantom{2} &\phantom{2} 3 \phantom{2}& \phantom{2}4 \phantom{2}\\
\hline
    $a$&  6+2& $(0,1)\times (1,-1)\times (1,-1)$ & 0 & 0  & 1  & 2 & -1 & 1 & 0 & 0 \\
    $b$&  4& $(-1,1)\times (0,1)\times (-1,3)$ & -2 & 2  &-  &- & -3 & 0 & 1 & 0   \\
\hline
    1&   8& $(1,0)\times (1,0)\times (2,0)$ & \multicolumn{8}{c|}{$3x_A=3x_B=x_C$}\\
    2&   2& $(1,0)\times (0,1)\times (0,-2)$& \multicolumn{8}{c|}{}\\
    3&   4& $(0,1)\times (1,0)\times (0,-2)$ & \multicolumn{8}{c|}{}\\
    4&   2& $(0,1)\times (0,-1)\times (2,0)$ & \multicolumn{8}{c|}{}\\
\hline
\end{tabular}
%\end{center}
%\caption{Model I.3} \label{modelI.3}
%\end{table}

%\begin{table}
%[htb] \footnotesize
%\renewcommand{\arraystretch}{1.0}
%\begin{center}
\begin{tabular}{|c||c|c||c|c|c|c|c|c|c|}
\hline
    \rm{model} I.4 & \multicolumn{9}{c|}{\phantom{morespac}$U(3)\times U(1)\times U(2)\times USp(10)\times USp(4)\times USp(2)$\phantom{morespac}}\\
\hline\hline \rm{stack} & $N$ & $(n^1,l^1)\times (n^2,l^2)\times
(n^3,l^3)$ &\phantom{2} $n_{\Ysymm}$ \phantom{2}&\phantom{2} $n_{\Yasymm}$ \phantom{2} & \phantom{2 }$b$\phantom{2 } & \phantom{2 }$b'$ \phantom{2 }& \phantom{2 }1\phantom{2 } & \phantom{2 }2\phantom{2 } &\phantom{2 } 3\phantom{2 }\\
\hline
    $a$&  6+2& $(0,1)\times (1,-1)\times (1,-1)$ & 0 & 0  & 0  & 3 & -1 & 1 & 0 \\
    $b$&   4& $(1,-1)\times (-1,2)\times (1,-1)$ & -2 & -6  &-  &- & -2& 1 & 2\\
\hline
    1&   10& $(1,0)\times (1,0)\times (2,0)$ & \multicolumn{7}{c|}{$x_A=x_B>2x_C$}\\
    2&   4& $(1,0)\times (0,1)\times (0,-2)$& \multicolumn{7}{c|}{}\\
    3&   2& $(0,1)\times (1,0)\times (0,-2)$ & \multicolumn{7}{c|}{}\\
\hline
\end{tabular}
\end{center}
\caption{D6-brane configurations and intersection numbers for the
three-family Standard-like models. Continued from table
\ref{models}.} \label{models2}
\end{table}

 We first consider three-family models with gauge group $U(4)\times U(2)\times
 U(1)$. Here $SU(4)$ subgroup of  $U(4)$ corresponds
to the Pati-Salam symmetry, $SU(2)$ subgroup  of $U(2)$
corresponds to the $SU(2)_L$ of the Standard-model  symmetry and
$U(1)$ is the additional non-anomalous gauge symmetry associated
with the  Cartan  generator of the $SU(2)_R$ symmetry of the
left-right symmetric model. Namely, the obtained  the Standard
Model gauge group structure is a descendant of  the left-right
symmetric gauge group. Starting with a stack of 8 branes not
parallel to the orientifold planes (these may be split into two
parallel but not overlapping stacks of 6 and 2 branes giving
$U(3)$ and  $U(1)$ gauge groups respectively) a second stack of 4
branes is introduced to generate an additional $U(2)$ (Note that
the $SU(3)$ subgroup of  $U(3)$ is associated with $SU(3)_{color}$
of the Standard-model). The number of families of quarks and
leptons is given by the intersections of the $U(3)$ stack with the
$U(2)$ stack and its $\Omega R$-image. As mentioned in \cite{CSU2}
the requirement of three families forces at least one of the tori
to be tilted. Since all left handed fermions should transform
under the same $U(3)$ representation the intersections of the
$U(3)$ stack with the $U(2)$ stack and its image (in this order!)
must have the same sign.

Apparently the tadpole and supersymmetry conditions for this type
of models are less constraining than  for the three-family $SU(5)$
GUT models in \cite{CPS}. Although this makes it easier to find
consistent Standard-like models it seems more difficult to find
all possible models and show there are no more models of a certain
type as was done for GUT models. So, instead, in table
\ref{models} we present some examples and compare them with the
three-family model of \cite{CSU2}. In fact, model I.3 is precisely
this earlier model but with slightly different conventions (the
wrapping numbers have been assigned differently to the three
two-tori). All four models share a similar gauge group and
spectrum. The chiral spectra in the open string sector are
tabulated in Tables \ref{spectrumI.1}, \ref{spectrumI.2},
\ref{spectrumI.3}.

\begin{table}
[htb] \footnotesize
\renewcommand{\arraystretch}{1.0}
\begin{center}
\begin{tabular}{|c||c||c|c|c|c|c||c|c|c|}\hline
I.1 & $SU(3)\times SU(2)\times USp(2)\times USp(4)$ & $Q_3$ &
$Q_1$ &
$Q_2$ & $Q_8$ & $Q_8'$ & $Q_Y$ & $Q_8-Q_8'$ & Field \\
\hline\hline $ab'$ & $3 \times (3,2,1,1)$ &
1 & 0 & 1 & 0 & 0 & $\frac 16$ & 0 & $Q_L$\\
  & $3 \times (1,2,1,1)$ &
0 & 1 & 1 & 0 & 0 & $-\frac 12$ & 0 & $ L$\\
$a1$ & $2\times(\overline{3},1,1,1)$ &
$-1$ & 0 & 0 & $\pm 1$ & 0 & $\frac 13, -\frac 23$ & $\pm 1$ & $\overline{U}$, $\overline{D}$\\
     & $2\times(\overline{3},1,1,1)$ &
$-1$ & 0 & 0 & 0 & $\pm 1$ & $\frac 13, -\frac 23$ & $\mp 1$ & $\overline{U}$, $\overline{D}$\\
     & $2\times(1,1,1,1)$ &
0 & $-1$ & 0 & $\pm 1$ & 0 &  $1,0$ & $\pm 1$ & $\overline{E}$, $\overline{N}$\\
    & $2\times(1,1,1,1)$ &
0 & $-1$ & 0 & 0 & $\pm 1$ & $1,0$ & $\mp 1$ & $\overline{E}$, $\overline{N}$\\
$a2$ & $(3,1,2,1)$ & 1 & 0 & 0 & 0 & 0 & $\frac 16$ & 0 & \\
     & $(1,1,2,1)$ & 0 & 1 & 0 & 0 & 0 & $\frac 16$ & 0 &  \\
$b1$ & $2\times 2\times(1,\overline{2},1,1)$ & 0 & 0 & $-1$ & $\pm 1$ & 0 & $\pm \frac 12$ & $\pm 1$ & $H_U,\; H_D$\\
     & $2\times 2\times(1,\overline{2},1,1)$ & 0 & 0 & $-1$ & 0 & $\pm 1$ & $\pm \frac 12$ & $\mp 1$ & $H_U,\; H_D$\\
$b3$ & $3\times(1,2,1,4)$ & 0 & 0 & 1 & 0 & 0 & 0 & 0 & \\
$b_{\Ysymm}$ & $(1,3,1,1)$ & 0 & 0 & 2 & 0 & 0 & 0 & 0 & \\
$b_{\Yasymm}$ & $(1,1,1,1)$ & 0 & 0 & $-2$ & 0 & 0 & 0 & 0 & \\
\hline
\end{tabular}
\end{center}
\caption{The chiral spectrum in the open string sector of model
I.1. All three families of left-handed quarks and leptons appear
at the same intersection. There are eight Higgs doublets of each
type in this model. } \label{spectrumI.1}
\end{table}

\begin{table}
[htb] \footnotesize
\renewcommand{\arraystretch}{1.0}
\begin{center}
\begin{tabular}{|c||c||c|c|c|c|c||c|c|c|}\hline
I.2 & $SU(3)\times SU(2)\times USp(2)\times USp(4)$ & $Q_3$ &
$Q_1$ &
$Q_2$ & $Q_8$ & $Q_8'$ & $Q_Y$ & $Q_8-Q_8'$ & Field \\
\hline\hline $ab$ & $(3,\overline{2},1,1)$ &
1 & 0 & $-1$ & 0 & 0 & $\frac 16$ & 0 & $Q_L$\\
  & $(1,\overline{2},1,1)$ &
0 & 1 & $-1$ & 0 & 0 & $-\frac 12$ & 0 & $ L$\\
$ab'$ & $2\times (3,2,1,1)$ &
1 & 0 & 1 & 0 & 0 & $\frac 16$ & 0 & $Q_L$\\
  & $2 \times (1,2,1,1)$ &
0 & 1 & 1 & 0 & 0 & $-\frac 12$ & 0 & $ L$\\
$a1$ & $2\times(\overline{3},1,1,1)$ &
$-1$ & 0 & 0 & $\pm 1$ & 0 & $\frac 13, -\frac 23$ & $\pm 1$ & $\overline{U}$, $\overline{D}$\\
     & $2\times(\overline{3},1,1,1)$ &
$-1$ & 0 & 0 & 0 & $\pm 1$ & $\frac 13, -\frac 23$ & $\mp 1$ & $\overline{U}$, $\overline{D}$\\
     & $2\times(1,1,1,1)$ &
0 & $-1$ & 0 & $\pm 1$ & 0 & $1, 0$ & $\pm 1$ & $\overline{E}$, $\overline{N}$\\
    & $2\times(1,1,1,1)$ &
0 & $-1$ & 0 & 0 & $\pm 1$ & $1, 0$ & $\mp 1$ & $\overline{E}$, $\overline{N}$\\
$a2$ & $(3,1,2,1)$ & 1 & 0 & 0 & 0 & 0 & $\frac 16$ & 0 & \\
     & $(1,1,2,1)$ & 0 & 1 & 0 & 0 & 0 & $\frac 16$ & 0 &  \\
$b1$ & $2\times 6\times(1,\overline{2},1,1)$ & 0 & 0 & $-1$ & $\pm 1$ & 0 & $\pm \frac 12$ & $\pm 1$ & $H_U,\; H_D$\\
     & $2\times 6\times(1,\overline{2},1,1)$ & 0 & 0 & $-1$ & 0 & $\pm 1$ & $\pm \frac 12$ & $\mp 1$ & $H_U,\; H_D$\\
$b3$ & $(1,2,1,4)$ & 0 & 0 & 1 & 0 & 0 & 0 & 0 & \\
$b_{\Ysymm}$ & $5\times(1,\overline{3},1,1)$ & 0 & 0 & -2 & 0 & 0 & 0 & 0 & \\
$b_{\Yasymm}$ & $5\times(1,1,1,1)$ & 0 & 0 & $2$ & 0 & 0 & 0 & 0 & \\
\hline
\end{tabular}
\end{center}
\caption{The chiral spectrum of model I.2. The model contains 24
Higgs doublets of each type.} \label{spectrumI.2}
\end{table}

\begin{table}
[htb] \footnotesize
\renewcommand{\arraystretch}{1.0}
\begin{center}
\begin{tabular}{|c||c||c|c|c|c|c||c|c|c|}\hline
I.3 & $SU(3)\times SU(2)\times USp(2)_2\times USp(4)\times
USp(2)_4$ & $Q_3$ & $Q_1$ &
$Q_2$ & $Q_8$ & $Q_8'$ & $Q_Y$ & $Q_8-Q_8'$ & Field \\
\hline\hline $ab$ & $(3,\overline{2},1,1,1)$ &
1 & 0 & $-1$ & 0 & 0 & $\frac 16$ & 0 & $Q_L$\\
  & $(1,\overline{2},1,1,1)$ &
0 & 1 & $-1$ & 0 & 0 & $-\frac 12$ & 0 & $ L$\\
$ab'$ & $2\times (3,2,1,1)$ &
1 & 0 & 1 & 0 & 0 & $\frac 16$ & 0 & $Q_L$\\
  & $2 \times (1,2,1,1,1)$ &
0 & 1 & 1 & 0 & 0 & $-\frac 12$ & 0 & $ L$\\
$a1$ & $2\times(\overline{3},1,1,1,1)$ &
$-1$ & 0 & 0 & $\pm 1$ & 0 & $\frac 13, -\frac 23$ & $\pm 1$ & $\overline{U}$, $\overline{D}$\\
     & $2\times(\overline{3},1,1,1,1)$ &
$-1$ & 0 & 0 & 0 & $\pm 1$ & $\frac 13, -\frac 23$ & $\mp 1$ & $\overline{U}$, $\overline{D}$\\
     & $2\times(1,1,1,1,1)$ &
0 & $-1$ & 0 & $\pm 1$ & 0 & $1, 0$ & $\pm 1$ & $\overline{E}$, $\overline{N}$\\
    & $2\times(1,1,1,1,1)$ &
0 & $-1$ & 0 & 0 & $\pm 1$ & $1,0$ & $\mp 1$ & $\overline{E}$, $\overline{N}$\\
$a2$ & $(3,1,2,1,1)$ & 1 & 0 & 0 & 0 & 0 & $\frac 16$ & 0 & \\
     & $(1,1,2,1,1)$ & 0 & 1 & 0 & 0 & 0 & $\frac 16$ & 0 &  \\
$b1$ & $2\times 3\times(1,\overline{2},1,1,1)$ & 0 & 0 & $-1$ & $\pm 1$ & 0 & $\pm \frac 12$ & $\pm 1$ & $H_U,\; H_D$\\
     & $2\times 3\times(1,\overline{2},1,1,1)$ & 0 & 0 & $-1$ & 0 & $\pm 1$ & $\pm \frac 12$ & $\mp 1$ & $H_U,\; H_D$\\
$b3$ & $(1,2,1,4,1)$ & 0 & 0 & 1 & 0 & 0 & 0 & 0 & \\
$b_{\Ysymm}$ & $2\times(1,\overline{3},1,1,1)$ & 0 & 0 & -2 & 0 & 0 & 0 & 0 & \\
$b_{\Yasymm}$ & $2\times(1,1,1,1,1)$ & 0 & 0 & $2$ & 0 & 0 & 0 & 0 & \\
\hline
\end{tabular}
\end{center}
\caption{The chiral spectrum of model I.3. This is precisely the
first three-family supersymmetric model obtained in
\cite{CSU1,CSU2,CSU3}. } \label{spectrumI.3}
\end{table}

\begin{table}
[htb] \footnotesize
\renewcommand{\arraystretch}{1.0}
\begin{center}
\begin{tabular}{|c||c||c|c|c|c|c|c||c|c|c|}\hline
I.4 & $SU(3)\times SU(2)\times USp(4)\times USp(2)$ & $Q_3$ &
$Q_1$ &
$Q_2$ & $Q_8$ & $Q_8'$ & $Q_{10}$& $Q_Y$ & $Q_8-Q_8'$ & Field \\
\hline\hline $ab'$ & $3 \times (3,2,1,1)$ &
1 & 0 & 1 & 0 & 0 & 0 & $\frac 16$ & 0 & $Q_L$\\
  & $3 \times (1,2,1,1)$ &
0 & 1 & 1 & 0 & 0 & 0 & $-\frac 12$ & 0 & $ L$\\
$a1$ & $2\times(\overline{3},1,1,1)$ &
$-1$ & 0 & 0 &  $\pm 1$ & 0 & 0 & $\frac 13, -\frac 23$ & $\pm 1$ & $\overline{U}$, $\overline{D}$\\
     & $2\times(\overline{3},1,1,1)$ &
$-1$ & 0 & 0 & 0 & $\pm 1$ & 0 & $\frac 13, -\frac 23$ & $\mp 1$ & $\overline{U}$, $\overline{D}$\\
    & $(\overline{3},1,1,1)$ &
$-1$ & 0 & 0 & 0 & 0 & $\pm 1$ & $\frac 13, -\frac 23$ & 0 &  $\overline{U}$, $\overline{D}$\\
     & $2\times(1,1,1,1)$ &
0 & $-1$ & 0 & $\pm 1$ & 0 & 0 & $1,0$ & $\pm 1$ & $\overline{E}$, $\overline{N}$\\
    & $2\times(1,1,1,1)$ &
0 & $-1$ & 0 & 0 &  $\pm 1$ & 0 & $1,0$ & $\mp 1$ & $\overline{E}$, $\overline{N}$\\
    & $(1,1,1,1)$ &
0 & $-1$ & 0 & 0 &  0 & $\pm 1$ & $1,0$ & 0 & $\overline{E}$, $\overline{N}$\\
$a2$ & $(3,1,4,1)$ & 1 & 0 & 0 & 0 & 0 & 0& $\frac 16$ & 0 & \\
     & $(1,1,4,1)$ & 0 & 1 & 0 & 0 & 0 & 0& $\frac 16$ & 0 &  \\
$b1$ & $2\times 2\times(1,\overline{2},1,1)$ & 0 & 0 & $-1$ &
    $\pm 1$ & 0 & 0& $\pm \frac 12$ & $\pm 1$ & $H_U,\; H_D$\\
     & $2\times 2\times(1,\overline{2},1,1)$ & 0 & 0 & $-1$ & 0 &  $\pm 1$ & 0& $\pm \frac 12$ & $\mp 1$ & $H_U,\; H_D$\\
     & $2\times(1,\overline{2},1,1)$ & 0 & 0 & $-1$ & 0 & 0 & $\pm 1$ & $\pm \frac 12$ & 0 & $H_U,\; H_D$\\
$b3$ & $2\times(1,2,1,2)$ & 0 & 0 & 1 & 0 & 0 & 0 & 0 & 0 & \\
$b_{\Ysymm}$ & $2\times(1,\overline{3},1,1)$ & 0 & 0 & $-2$ & 0 & 0 &0& 0 & 0 & \\
$b_{\Yasymm}$ & $6\times(1,1,1,1)$ & 0 & 0 & $-2$ & 0 & 0 &0 & 0 & 0 & \\
\hline
\end{tabular}
\end{center}
\caption{The chiral spectrum of model I.4. As for model I.1 all
three families of left-handed quarks and leptons come from the
same intersection but here there are twelve Higgs doublets of each
type as in the original model I.3.} \label{spectrumI.4}
\end{table}

To generate right-handed quarks and leptons (and of course the
appropriate assignments for the
 $U(1)$ hypercharge) we break the $USp(8)$
or $USp(10)$ factor to a number of $U(1)$s by moving the branes
away from the orientifold planes. From the action of the
orientifold group \ref{orbifold} and \ref{orientifold} we see that
a D6-brane which is away from the orientifold plane in one of the
three 2-tori has one image under the orientifold group. Similarly,
a brane away from the orientifold plane in two or three tori has
two and three images respectively. This means that we can break
$USp(2)$ to a single $U(1)$ by moving the 3-cycle away from the
orientifold plane in one torus (the original two branes on top of
the orientifold plane account for a single brane away from the
orientifold plane and its image). $USp(4)$ can be broken to either
two $U(1)$s through two $USp(2)$s or down to a single $U(1)$ by
moving the brane away from the orientifold plane in all three tori
(thus generating four images). For $USp(6)$ we can have the
following breaking patterns \beqa USp(6)\rightarrow USp(2)\times
USp(2)\times USp(2)\rightarrow U(1)\times U(1)\times U(1)\\ \NO
USp(6)\rightarrow USp(4)\times USp(2)\rightarrow\ldots \\\NO
USp(6)\rightarrow U(1)\times U(1) \eeqa where the last scheme
arises when two of the original six branes are displaced away from
the orientifold plane in two tori, thus each gaining two images
under the orientifold group. Since the $U(1)$s obtained this way
are non-anomalous they do not acquire string scale mass and so
extend the gauge group beyond the Standard Model. To avoid having
too many such factors surviving we break $USp(8)$ and $USp(10)$ as
\beqa
USp(8)\rightarrow USp(4)\times USp(4)\rightarrow U(1)\times U(1)\\
\NO USp(10)\rightarrow USp(8)\times USp(2)\rightarrow U(1)\times
U(1)\times U(1)\eeqa resulting in the minimal number of $U(1)$s.

If we denote by $Q_1$, $Q_2$ and $Q_3$ the $U(1)$ factor of the
corresponding $U(n)$, the linear combination $Q_3/3-Q_1=B-L$ is
non-anomalous because it corresponds to a generator in $SU(4)$
when all eight branes are coincident. Here $B=Q_3$ is the baryon
number and $L=Q_1$ plays the role of lepton number. In models I.1,
I.2 and I.3 there are three non anomalous $U(1)$ factors, namely
$B-L$, $Q_8$ and $Q_8'$, the last two coming from breaking
$USp(8)$, and two anomalous, $B+L$ and $Q_2$. In model I.4 there
is an extra non-anomalous $U(1)$, $Q_{10}$, coming from the
$USp(2)$ of $USp(10)$. The hypercharge is defined as the
non-anomalous linear combination \beq Q_Y=\frac 16 Q_3-\frac 12
Q_1+\frac 12 (Q_8+Q_8'+Q_{10}). \eeq

The phenomenology and the tree-level gauge couplings of the
three-family model in \cite{CSU2} were discussed in detail in
\cite{CLS1}.  Yukawa couplings in this context were discussed in
\cite{CLS2} and more recently studied in detail in
\cite{CIM,CP,AO}. The models we present here have very similar
properties and so we will simply point out the essential
differences.

A common feature of the new models is that they have less filler
branes than the original model I.3. In fact model I.4 is the first
example that includes a stack with no zero angles relative to the
$\Omega R$ plane, and hence there is an additional complex
structure modulus, that is not fixed. In fact there is a
two-parameter family of such models in moduli space. Models I.1
and I.4 give three families of left-handed quarks and leptons from
the same intersection, in contrast to the rest of the models where
two families are obtained from one intersection and the third from
another. The original model had 24 Higgs doublets (12 $H_U$ and 12
$H_D$). This is somewhat improved in model I.1 (8+8 Higgs
doublets) and I.4 (10+10 Higgs doublets) although model I.2
contains 24+24 Higgs doublets. Additionally, model I.4 has five
families of right handed quarks and leptons compared to four in
the rest of the models. In fact, since the right handed quarks and
leptons are obtained by breaking $USp(8)$ to two $U(1)$s, one can
get less families by less breaking of $USp(8)$ but then the
spectrum would contain more exotic multiplets. The existence of
extra non-anomalous $U(1)$s is a generic feature of these
supersymmetric models. Typically, one of these additional $U(1)$'s
can be broken by the scalar component of ${\bar N}$, i.e. the
right-handed neutrino super-multiplet.  However, for the
additional $U(1)$s there are no Higgs field candidates to
spontaneously  break them (see \cite{CLS1} for a discussion).

\subsection{A left-right symmetric model}

Here we present an example of a supersymmetric left-right
symmetric model with three generations of quarks and leptons. The
brane configuration is shown in Table \ref{modelII} and the chiral
spectrum in the open string sector is tabulated in Table
\ref{spectrumII}. The electromagnetic charge is obtained as \beq
Q_{em}=I_{3L}+I_{3R}+\frac 12(B-L), \eeq where $I_{3L}$, $I_{3R}$
are the Cartan generators of $SU(2)_L$ and $SU(2)_R$,
respectively, and $B-L$ is identified as before with the $U(1)$ in
the breaking of $SU(4)\rightarrow SU(3)\times U(1)_{B-L}$.

The Higgs sector of the  model , however, contains 16 Higgs
bi-doublets. In addition, the symmetric multiplets coming from
$SU(2)_L$ and $SU(2)_R$ are not charged under $B-L$ and hence they
cannot be used to break $SU(2)_R$. Nevertheless, $SU(2)_R$ could
be broken to $U(1)$ by separating the two branes but, again, there
is no mechanism to break this non-anomalous $U(1)$.

\begin{table}
[htb] \footnotesize
\renewcommand{\arraystretch}{1.0}
\begin{center}
\begin{tabular}{|c||c|c||c|c|c|c|c|c|c|c|}
\hline
    \rm{model} II & \multicolumn{10}{c|}{\phantom{more space}$U(4)\times U(2)_R\times U(2)_L\times USp(2)\times USp(2)$\phantom{more space}}\\
\hline\hline \rm{stack} & $N$ & $(n^1,l^1)\times (n^2,l^2)\times
(n^3,l^3)$ & \phantom{2}$n_{\Ysymm}$\phantom{2} & \phantom{2}$n_{\Yasymm}$\phantom{2}  & \phantom{2 }$b$\phantom{2 } & \phantom{2 }$b'$\phantom{2 } & \phantom{2 }$c$ \phantom{2 }& \phantom{2 }$c'$\phantom{2 } & \phantom{2 }2 \phantom{2 }& \phantom{2 }4\phantom{2 }\\
\hline
    $a$&  8& $(-1,1)\times (-1,0)\times (1,1)$ & 0 & 0  & -1  & -2 & 1 & 2 & -1 & 1\\
    $b$&  4& $(-1,0)\times (-2,1)\times (1,3)$ & 5 & -5  &-  &- & 16 & 0 & 0 & 6\\
    $c$&  4& $(0,-1)\times (2,1)\times (3,1)$ & -5 & 5  &-  &- & -& - & -6 & 0\\
\hline
    2&   2& $(1,0)\times (0,1)\times (0,-2)$& \multicolumn{8}{c|}{$x_A=x_C=6x_B$}\\
    4&   2& $(0,1)\times (0,-1)\times (2,0)$ & \multicolumn{8}{c|}{}\\
\hline
\end{tabular}
\end{center}
\caption{D6-brane configurations and intersection numbers for the
three-family left-right symmetric model} \label{modelII}
\end{table}

\begin{table}
[htb] \footnotesize
\renewcommand{\arraystretch}{1.0}
\begin{center}
\begin{tabular}{|c||c||c|c|c||c|c|c|}\hline
II & $SU(4)\times SU(2)_R\times SU(2)_L\times USp(2)\times USp(2)$
& $Q_4$ & $Q_2$ & $Q_2'$ & $Q_{em}$ & $B-L$ & Field \\
\hline\hline
$ab$ & $(\overline{4},2,1,1,1)$ & $-1$ & 1 & 0  & $\frac 13,\; -\frac 23,\;1,\; 0$ & $-\frac 13,\;1$ & $Q_R, R$\\
$ab'$ & $2 \times (\overline{4},\overline{2},1,1,1)$ &
$-1$ & -1 & 0  & $\frac 13,\; -\frac 23,\;1,\; 0$ & $-\frac 13,\;1$ & $Q_R, R$ \\
$ac$ & $(4,1,\overline{2},1,1)$ & 1 & 0 & $-1$   & $\frac 23,\; -\frac 13,\;0,\; -1$ & $\frac 13,\;-1$ & $Q_L, L$\\
$ac'$ & $2 \times (4,1,2,1,1)$ & 1 & 0 & 1  & $\frac 23,\; -\frac 13,\;0,\; -1$ & $\frac 13,\;-1$ & $Q_L, L$ \\
$a2$ & $(\overline{4},1,1,2,1)$ & $-1$ & 0 & 0 & $-\frac 16,\;\frac 12$ & $-\frac 13,\;1$ & \\
$a4$ & $(4,1,1,1,2)$ & 1 & 0 & 0   & $\frac 16,\;-\frac 12$ & $\frac 13,\;-1$ & \\
$bc$ & $16\times(1,2,\overline{2},1,1)$ & 0 & 1 & $-1$   & $0,\;0,\;\pm 1$ & 0 & $H$\\
$b4$ & $6\times(1,2,1,1,2)$ & 0 & 1 & 0   & $\pm \frac 12$ & 0 & \\
$c2$ & $6\times(1,1,\overline{2},2,1)$ & 0 & 0 & -1   & $\pm \frac 12$ & 0 & \\
$b_{\Ysymm}$ & $5\times(1,3,1,1,1)$ & 0 & 2 & 0   & $0,\pm 1$ & 0 & \\
$b_{\Yasymm}$ & $5\times(1,1,1,1,1)$ & 0 & -2 & 0   & 0 & 0 & \\
$c_{\Ysymm}$ & $5\times(1,1,\overline{3},1,1)$ & 0 & 0 & -2   & $0,\pm 1$ & 0 & \\
$c_{\Yasymm}$ & $5\times(1,1,1,1,1)$ & 0 & 0 & 2   & 0 & 0 & \\
\hline
\end{tabular}
\end{center}
\caption{The chiral spectrum in the open string sector of model
II} \label{spectrumII}
\end{table}

\section{Conclusions}
By generalizing the original supersymmetric constructions within $
{\IT^6/( \IZ_2\times \IZ_2)}$ orientifolds with intersecting
D6-branes we presented new supersymmetric Standard-like  models.
In particular we allowed for D6-branes wrapping most general
three-cycles, which are product of  one-cycles on each two-torus,
and allowing for one two-torus to be tilted.  As a consequence we
obtained new models with the three families of quarks and leptons
and the Standard-model gauge group as a part of the gauge
structure. Some of the models have fewer   Higgs doublets
(however, still at least 8 $H_U$ and 8 $H_D$ Higgs doublets). We
have also found models that  have all the three  families of
left-handed quarks (and leptons) arising from the same
intersection sector. This feature may allow potentially for a new
pattern  of Yukawa couplings of quarks and leptons to Higgs
doublets. (In the original model \cite{CSU1,CSU2} one family came
from a different sector and as a consequence did not have Yukawa
couplings to the Higgs doublets.)

We also obtained a genuinely left-right symmetric model with three
copies of left-handed {\it  and }  right-handed quarks and
leptons. Nevertheless, this model still has a large number (16) of
Higgs bi-doublets, as well as chiral matter charged under the
left-right symmetric gauge group and the additional gauge sector.

In these models, the origin of the hypercharge is within
left-right symmetric (Pati-Salam) gauge  group structure, i.e. the
non-anomalous hypercharge is obtained as a linear combination of
non-anomalous U(1) charges that are a part of  the  Pati-Salam
gauge group \cite{PatiSalam}. It would be interesting to
generalize these constructions to cases where the hypercharge
would not be a descendant of the left-right symmetric model. We
hope to address these types of models in the future.

The models presented in this paper all have additional,
semi-hidden, gauge group structure that is typically confining,
i.e. the additional gauge group factors have negative beta
functions.  This sector may therefore allow for a possibility of
dynamical supersymmetry breaking and moduli stabilization as
addressed for the original model \cite{CSU1,CSU2} in \cite{CLW}.
It would be very interesting to address dynamical supersymmetry
breaking for the new  classes of Standard-like models, presented
in this paper.

\section*{Acknowledgments}

We would like to thank    Paul Langacker and Gary
Shiu  for useful discussions. M.C. would like to
thank the New Center for Theoretical Physics at Rutgers University
and the Institute for Advanced Study, Princeton,
 for hospitality and
support  during the course of this work. Research supported in
part by DOE grant DOE-FG02-95ER40893, NATO linkage grant No. 97061
(M.C.) and Fay R. and Eugene L. Langberg Chair (M.C.).


\begin{thebibliography}{99}

\bibitem{bgkl}
R.~Blumenhagen, L.~G\"orlich, B.~K\"ors and D.~L\"ust,
%`Noncommutative compactifications of type I strings on tori with magnetic
%background flux',
JHEP {\bf 0010} (2000) 006.

\bibitem{afiru}
G.~Aldazabal, S.~Franco, L.~E.~Ib\'a\~nez, R.~Rabad\'an and
A.~M.~Uranga,
%`D = 4 chiral string compactifications from intersecting branes'
Journal of Mathematical Physics, vol. 42, number 7, p. 3103, {
hep-th/0011073}; JHEP {\bf 0102} (2001) 047.

\bibitem{bkl}
R.~Blumenhagen, B.~K\"ors and D.~L\"ust,
%`Type I strings with F flux and B flux',
JHEP {\bf 0102} (2001) 030.

\bibitem{imr}
L.~E.~Ib\'a\~nez, F.~Marchesano and R.~Rabad\'an,
%`Getting just the standard model at intersecting branes'
{ hep-th/0105155}.

\bibitem{magnetised}
C.~Angelantonj, I.~Antoniadis, E.~Dudas and A.~Sagnotti,
%`Type I strings on magnetized orbifolds and brane transmutation',
Phys. Lett. B {\bf 489} (2000) 223.

\bibitem{bdl}
M.~Berkooz, M.~R.~Douglas and R.~G.~Leigh,
%`Branes intersecting at angles'
Nucl. Phys. B {\bf 480} (1996) 265.

\bibitem{bachas}
C.~Bachas,
%``A Way  to  break  supersymmetry,''
%arXiv:
hep-th/9503030.

%%%%%%%%%%%%%%%%%%%%%%%%%%%%%%%%%%%%%%%%%%%%%%%%%%%%%%%%%%%%%%%%%%%%%%%%%
%% subsequent work on branes at angles
%%%%%%%%%%%%%%%%%%%%%%%%%%%%%%%%%%%%%%%%%%%%%%%%%%%%%%%%%%%%%%%%%%%%%%%%%

\bibitem{bonn}
S.~F\"orste, G.~Honecker and R.~Schreyer,
%`Supersymmetric $\IZ_N \times \IZ_M$ orientifolds in 4-D with D branes at
%angles',
Nucl. Phys. B {\bf 593} (2001) 127; JHEP {\bf 0106} (2001) 004.

\bibitem{bklo}
R.~Blumenhagen, B.~K\"ors and D.~L\"ust, T.~Ott,
%`Noncommutative compactifications of type I strings on tori with magnetic
%background flux',
Nucl. Phys. {\bf B616} (2001) 3.

\bibitem{bailin}    D.~Bailin, G.~V.~Kraniotis, and A.~Love, Phys.\ Lett.\
B {\bf 530}, 202 (2002);
%[arXiv:hep-th/0108131];
Phys.\ Lett.\ B {\bf 547}, 43 (2002);
%[arXiv:hep-th/0208103];
%arXiv:
hep-th/0210219;
%arXiv:
hep-th/0212112.

%%CITATION = HEP-TH 0011289;%%Bailin et al.
\bibitem{kokorelis}  C.~Kokorelis,
%``New standard model vacua from intersecting branes,''
JHEP {\bf 0209}, 029 (2002);
%[arXiv:hep-th/0205147];
JHEP {\bf 0208}, 036 (2002);
%[arXiv:hep-th/0206108];
        hep-th/0207234; JHEP {\bf 0211}, 027
(2002); hep-th/0210200.

\bibitem{CSU1}
M.~Cveti\v c, G.~Shiu and A.~M.~Uranga,
%``Three-family supersymmetric
%standard like models from intersecting  brane worlds,''
Phys.\ Rev.\ Lett.\  {\bf 87}, 201801 (2001).
% [arXiv:hep-th/0107143].
%%CITATION = HEP-TH 0107143;%%

\bibitem{CSU2}
M.~Cveti\v c, G.~Shiu and A.~M.~Uranga,
%``Chiral four-dimensional
%N = 1 supersymmetric type IIA orientifolds from  intersecting
%D6-branes,''
Nucl.\ Phys.\ B {\bf 615}, 3 (2001).
%[arXiv:hep-th/0107166]
%%CITATION = HEP-TH 0107166;%%

\bibitem{CSU3}
M.~Cveti\v c , G.~Shiu and A.~M.~Uranga,
%``Chiral type II orientifold constructions as M theory on G(2) holonomy
%space
%arXiv:
hep-th/0111179.
%%CITATION = HEP-TH 0111179;%%

\bibitem{AW}
M.~Atiyah and E.~Witten,
%``M-theory dynamics on a manifold of G(2) holonomy,''
hep-th/0107177.

\bibitem{Witten}
E.~Witten,
%``Anomaly cancellation on G(2)manifolds,''
hep-th/0108165.

\bibitem{aW}
B.~Acharya and E.~Witten,
%``Chiral fermions from manifolds of G(2) holonomy,''
hep-th/0109152.

\bibitem{FW}
T.~Friedmann and E.~Witten,
%``Unification scale, proton decay, and manifolds of G(2) holonomy,''
%arXiv:
hep-th/0211269.
%%CITATION = HEP-TH 0211269;%%

\bibitem{CPS} M. Cveti\v c, I. Papadimitriou and G. Shiu,
% ``A Three Family  SU(5) Grand Unified Models from Type IIA Orienfolds
%with Intersecting D6-Branes'',
 %arXiv:
hep-th/0212177, accepted for publication in Nucl. Phys. {\bf B}.

\bibitem{blumrecent}
R.~Blumenhagen, L.~Gorlich and T.~Ott,
%``Supersymmetric intersecting branes on the type IIA T**6/Z(4)  orientifold,''
%arXiv:
hep-th/0211059.
%%CITATION = HEP-TH 0211059;%%

\bibitem{Honecker}
G.~Honecker,
%``Chiral supersymmetric models on an orientifold of
%Z(4) x Z(2) with  intersec
%arXiv:
hep-th/0303015.
%%CITATION = HEP-TH 0303015;%%


%\bibitem{CIMD5}
%D.~Cremades, L.~E.~Ibanez and F.~Marchesano,
%``Standard model at intersecting D5-branes: Lowering the string scale,''
%Nucl.\ Phys.\ B {\bf 643}, 93 (2002) [arXiv:hep-th/0205074].
%%CITATION = HEP-TH 0205074;%%

%%%%%%%%%%%%%%%%%%%%%%%%%%%%%%%%%%%%%%%%%%%%%%%%%%%%%%%%%%%%%%%%%%%%%%%%
%% G2
%%%%%%%%%%%%%%%%%%%%%%%%%%%%%%%%%%%%%%%%%%%%%%%%%%%%%%%%%%%%%%%%%%%%%%%%

%\bibitem{acharya}
%B.~S.~Acharya,
%``On realising N = 1 super Yang-Mills in M theory,''
% hep-th/0011089.

%\bibitem{AMV}
%M.~Atiyah, J.~Maldacena and C.~Vafa,
%hep-th/0011256.

\bibitem{CLS1}
 M.~Cveti\v c, P.~Langacker and G.~Shiu,
%``Phenomenology of a three-family Standard-like string model,''
Phys.\ Rev.\ D {\bf 66}, 066004 (2002).
% [arXiv:hep-ph/0205252].
%%CITATION = HEP-PH 0205252;%%
 \bibitem{CLS2}
 M.~Cveti\v c, P.~Langacker and G.~Shiu,
%``A three-family Standard-like orientifold model: Yukawa couplings and
%hierarchy''
Nucl.\ Phys.\ B {\bf 642}, 139 (2002).
% [arXiv:hep-th/0206115].
%%CITATION = HEP-TH 0206115;%%

\bibitem{CLW}
M. Cveti\v c, P. Langacker and J. Wang, ``Dynamical Supersymmetry
Breaking and Standard-lime Models with Intersecting D6-branes'',
RUNHETC-2002-47, to appear.

\bibitem{CIM1}
D.~Cremades, L.~E.~Ibanez and F.~Marchesano,
%``Towards a theory of quark masses, mixings and CP-violation,''
%arXiv:
hep-ph/0212064.
%%CITATION = HEP-PH 0212064;%%
%\cite{Friedmann:2002ty}

\bibitem{Lust}
D.~L\"ust and S.~Stieberger,
%``Gauge threshold corrections in intersecting brane world models,''
%arXiv:
hep-th/0302221.
%%CITATION = HEP-TH 0302221;%%

%%%%%%%%%%%%%%%%%%%%%%%%%%%%%%%%%%%%%%%%%%%%%%%%%%%%%%%%
% Yiannis                                              %
%%%%%%%%%%%%%%%%%%%%%%%%%%%%%%%%%%%%%%%%%%%%%%%%%%%%%%%%
\bibitem{CIM}
D.~Cremades, L.~E.~Ibanez and F.~Marchesano,
%``Yukawa couplings in intersecting D-brane models,''
%arXiv:
hep-th/0302105.
%%CITATION = HEP-TH 0302105;%%
\bibitem{CP}
M.~Cveti\v c and I.~Papadimitriou,
%``Conformal field theory couplings for intersecting D-branes on
%orientifolds,''
%arXiv:
hep-th/0303083.
%%CITATION = HEP-TH 0303083;%%

\bibitem{AO}
S.~A.~Abel and A.~W.~Owen,
%``Interactions in Intersecting Brane Models,''
%arXiv:
hep-th/0303124.
%%CITATION = HEP-TH 0303124;%%

\bibitem{PatiSalam}
J.~C.~Pati and A.~Salam,
%``Unified Lepton - Hadron Symmetry And A Gauge Theory Of The
%Basic Interactions,''
Phys.\ Rev.\ D {\bf 8} (1973) 1240.
%%CITATION = PHRVA,D8,1240;%%

\end{thebibliography}
\end{document}